# The NIR Spectrograph for the new SOXS instrument at the NTT

F. Vitali*[a], M. Aliverti[b], G. Capasso[c], F. D'Alessio[a], M. Munari[d], M. Riva[b], S. Scuderi[d], R. Zanmar Sanchez[d], S. Campana[b], P. Schipani[c], R. Claudi[e], A. Baruffolo[e], S. Ben-Ami[f], F. Biondi[e], A. Brucalassi[g,m], R. Cosentino[h,d], D. Ricci[e], P. D'Avanzo[b], O. Diner[i], H. Kuncarayakti[j,k], A. Rubin[i], J. Achrén[l], J. A. Araiza-Duran[m,n], I. Arcavi[o], A. Bianco[b], E. Cappellaro[e], M. Colapietro[c], M. Della Valle[c], S. D'Orsi[c], D. Fantinel[e], J. Fynbo[p], A. Gal-Yam[i], M. Genoni[b], M. Hirvonen[q], J. Kotilainen[j,k], T. Kumar[k], M. Landoni[b], J. Lehti[q], G. Li Causi[r], L. Marafatto[e], S. Mattila[k], G. Pariani[b], G. Pignata[m,n], M. Rappaport[i], B. Salasnich[e], S. Smartt[s], M. Turatto[e]

[a]INAF–Osservatorio Astronomico di Roma, Via Frascati 33, I-00078 M. Porzio Catone, Italy; [b]INAF–Osservatorio Astronomico di Brera, Via Bianchi 46, I-23807, Merate, Italy; [c]INAF–Osservatorio Astronomico di Capodimonte, Sal. Moiariello 16, I-80131, Naples, Italy; [d]INAF–Osservatorio Astrofisico di Catania, Via S. Sofia 78, I-95123 Catania, Italy; [e]INAF–Osservatorio Astronomico di Padova, Vicolo dell'Osservatorio 5, I-35122, Padua, Italy; [f]Harvard-Smithsonian Center for Astrophysics, Cambridge, USA; [g]ESO, Karl Schwarzschild Strasse 2, D-85748, Garching bei München, Germany; [h]FGG-INAF, TNG, Rambla J.A. Fernandez Perez 7, E-38712 Brenã Baja (TF), Spain; [i]Weizmann Institute of Science, Herzl St 234, Rehovot, 7610001, Israel; [j]Finnish Centre for Astronomy with ESO (FINCA), FI-20014 University of Turku, Finland; [k]Tuorla Observatory, Dept. of Physics and Astronomy, FI-20014 University of Turku, Finland; [l]Incident Angle Oy, Capsiankatu 4 A 29, FI-20320 Turku, Finland; [m]Universidad Andres Bello, Avda. Republica 252, Santiago, Chile; [n]Millennium Institute of Astrophysics (MAS); [o]Tel Aviv University, Department of Astrophysics, 69978 Tel Aviv, Israel; [p]Dark Cosmology Centre, Juliane Maries Vej 30, DK-2100 Copenhagen, Denmark; [q]Aboa Space Research Oy, Tierankatu 4B, FI-20520 Turku, Finland; [r]INAF - Istituto di Astrofisica e Planetologia Spaziali, Via Fosso del Cavaliere 100, 00133, Roma, Italy; [s]Astrophysics Research Centre, Queen's University Belfast, Belfast, BT7 1NN, UK.

## Abstract

We present the NIR spectrograph of the Son Of XShooter (SOXS) instrument for the ESO-NTT telescope at La Silla (Chile). SOXS is a R~4,500 mean resolution spectrograph, with a simultaneously coverage from about 0.35 to 2.00 µm. It will be mounted at the Nasmyth focus of the NTT. The two UV-VIS-NIR wavelength ranges will be covered by two separated arms. The NIR spectrograph is a fully cryogenic echelle-dispersed spectrograph, working in the range 0.80-2.00 µm, equipped with an Hawaii H2RG IR array from Teledyne, working at 40 K. The spectrograph will be cooled down to about 150 K, to lower the thermal background, and equipped with a thermal filter to block any thermal radiation above 2.0 µm. In this poster we will show the main characteristics of the instrument along with the expected performances at the telescope.

*fabrizio.vitali@inaf.it

# 1. INTRODUCTION

Son Of X-Shooter (SOXS) will be the new UV-VIS-NIR spectrograph for the ESO 3.5m New Technology Telescope in La Silla (Chile). SOXS in an international project, leaded by the Italian National Institute of Astrophysics (INAF). The instrument will be mainly a spectroscopic facility for the follow-up of transient and variable sources. Recently, the instrument has successfully passed the Preliminary Design Review phase by ESO (July 2017) and is now in the Final Design Phase (July 2018). SOXS is a RS > 4,500 resolution spectrograph, with a simultaneously coverage from about 0.35 to 2.00 $\mu$m. Its design foresees two separate high-efficiency spectrographs (UV–VIS and NIR) with a small overlapping range of about 80 nm for the spectral intercalibration. The NIR spectrograph is a fully cryogenic echelle-dispersed spectrograph, working in the range 0.80-2.00 $\mu$m. The optical design is based on a 4C echelle. The dispersion is obtained via a main disperser grating and three prisms as cross-dispersers. The NIR spectrum will be dispersed on 15 orders, with a minimum inter-order of ~10 px (in the blue part). The average throughput will be about 28%, including the telescope. The spectrograph will be equipped with an Hawaii H2RG IR array from Teledyne, cooled down to 40 K and controlled via the new NGC controller from ESO. It will be cooled down to about 150 K, to lower the thermal background, and equipped with a thermal filter to block any thermal radiation above 2.00 $\mu$m. The cryogenics will be operated via a Closed Cycle CryoCooler.

# 2. THE NIR ARM OF SOXS

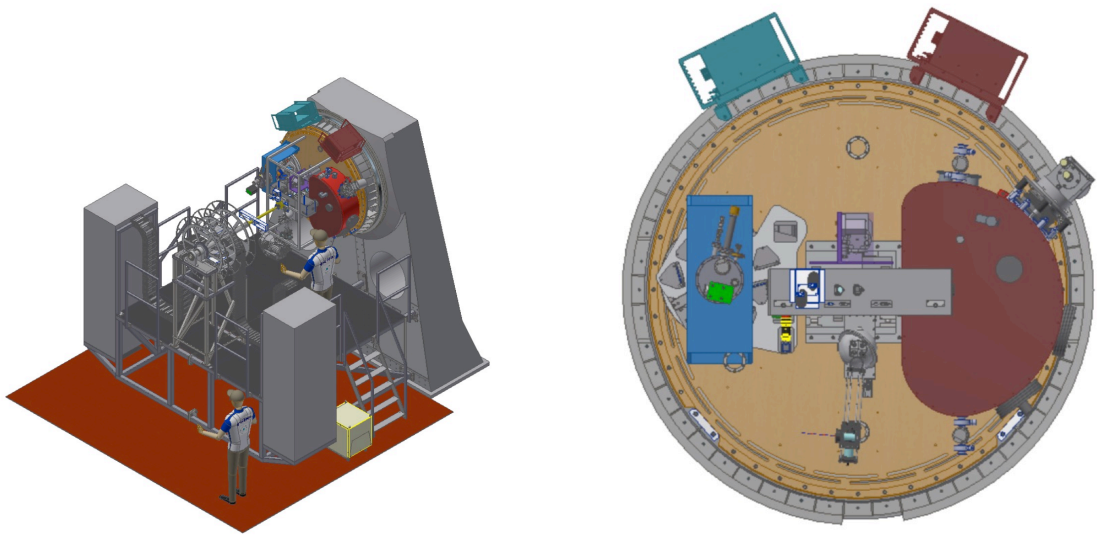

Figure 1. The whole SOXS instrument at the NTT Nasmyth focus (left) and the front view of the instrument at the derotator flange (right). The NIR Spectrograph is the D-shaped red cryostat on the right.

SOXS will see its first light at the end of 2020, mounted at the Nasmyth focus of NTT replacing SOFI [1][2]. The whole system (see **Figure 1**) is constituted by the three main scientific arms: the UV–VIS spectrograph, the NIR spectrograph and the acquisition camera. The three main arms, the calibration box and the NTT are connected together by the Common Path (CP) [3]. The NIR part of the CP is similar to the UV-VIS arm (same first order parameters, except the wavelength range). The NIR arm does not include an ADC, since the atmospheric dispersion is less severe than in the UV–VIS range and is considered acceptable. The CP NIR Arm (see **Figure 2**) includes a doublet (CP_LIR_01) in order to reduce the telescope F/11 beam to an F/6.5 beam.

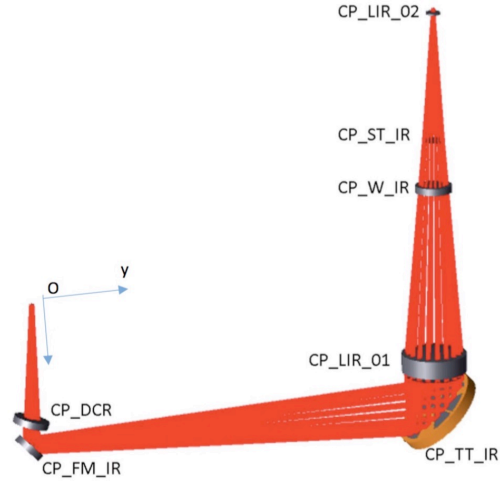

Figure 2. The layout of the NIR arm of the Common Path.

Two flat mirrors (CP_FM_IR and CP TT IR) relay the light to the slit. In order to allow the entering of light in the NIR spectrograph dewar, the CP NIR Arm includes a flat window (CP_W_IR), and, to reduce the noise in the the NIR spectrograph, a cold stop has been introduced after the window itself (CP_ST_IR). A field lens (CP LIR 02), placed near the slit, remaps the telescope pupil on the grating of the spectrograph.

## 3. THERMAL DESIGN

The SOXS NIR Spectrograph works in the range 0.80 - 2.00 μm, using the Hawaii H2RG Teledyne detector. In this wavelength range, the contribution of the continuum between the OH lines from the sky is very low and we have to take the thermal contribution of the spectrograph significantly less than that of the sky and the Dark Current. We decided to operate the detector at 40K, to avoid some excess of persistence showed by soma Hawaii array (see §6). At this temperature, the detector can show Dark Current level as low as $10^{-3}$ e$^-$·s. Then, we can apply Eq. 1 to derive the thermal flux from the spectrograph at a certain temperature.

Eq. 1 $$NTE = A \cdot \Omega \cdot \int \eta_{det}(\lambda) \cdot \varepsilon(\lambda) \cdot F(\lambda) \cdot 2c / (\lambda^4 \cdot \exp(hc/\lambda KT) - 1) \cdot d\lambda$$

where:
NTE = Number of Thermal electrons;
A = pixel area, 18x18 μm;
Ω = solid angle at pixel;

$\eta_{det}(\lambda)$ = detector efficiency (Q.E.);

$\varepsilon(\lambda)$ = atmosphere + telescope + instrument emissivity;
F ($\lambda$) = filter transmission;
$\lambda$ = wavelength, in μm.

In **Figure 3** the result for a temperature of 150 K is shown. This is a conservative value for the working temperature, to be sure that we will work in background limited conditions.

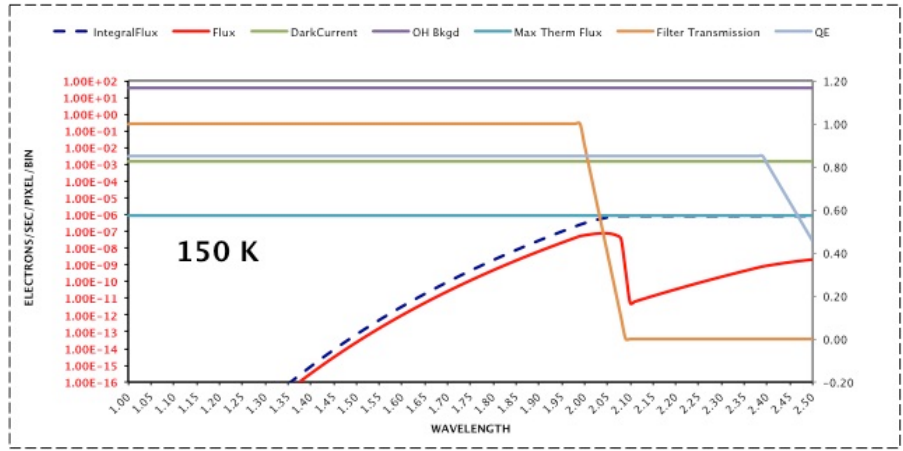

Figure 3. Thermal inputs from several sources at 150 K. The integrated thermal flux is well below the D.C. level.

## 4. THE OPTICAL DESIGN

The SOXS NIR spectrograph is a near infrared echelle spectrograph, with R=5000 (for 1 arcsec slit), covering a wavelength range from 800 to 2000 nm with 15 orders [4].

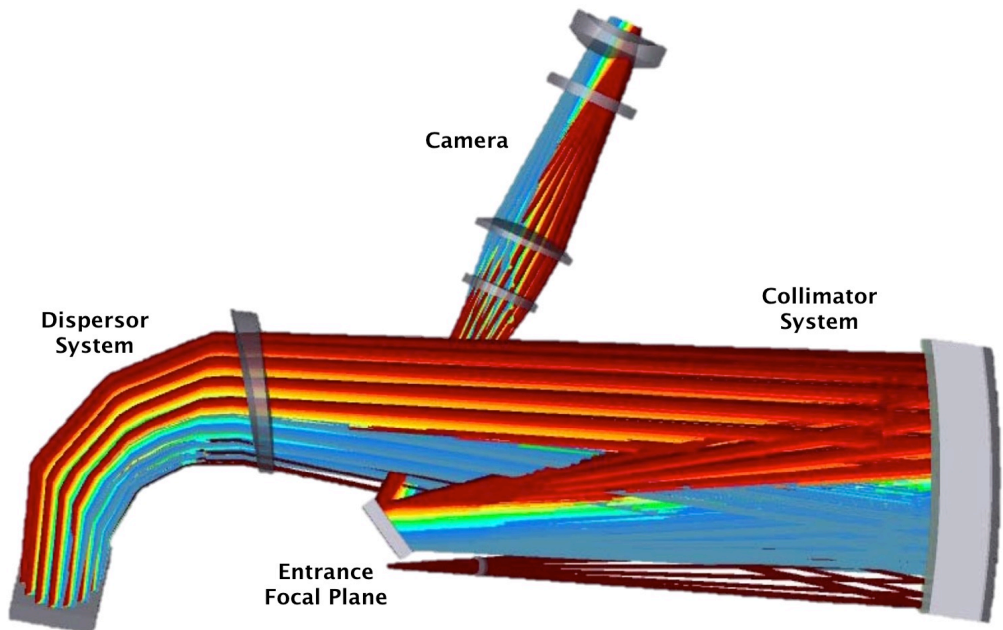

Figure 4. The layout of the NIR Spectrograph.

Based on the 4C layout (see Figure 4), the NIR spectrograph is composed of a double pass collimator and a refractive camera, a disperser (grating) and a cross disperser. The spectrograph shall collect the light coming from the CP, switching between different slit sizes (0.5", 1.0", 1.5", 5.0"). In addition to those slits, a pinhole (0.5") on the focal plane will be installed on the slit stage for alignment purpose.

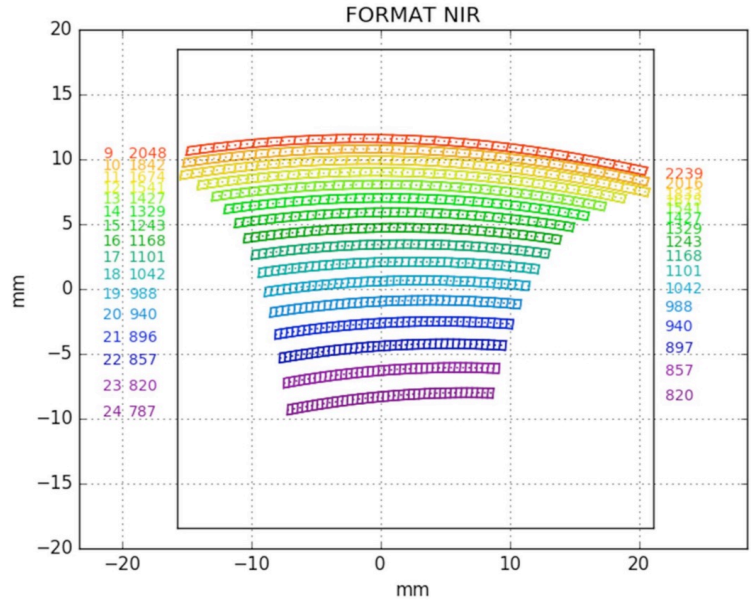

Figure 5. The echelle spectral format on the 2kx2k Hawaii array.

The collimator of the system is a Maksutov telescope, used off-axis (and consequently cut). The system has an input F/# of 6.5 (produced by the CP NIR arm) and produces a collimated beam of diameter ~50mm.

The main disperser is a standard grating, with 72 l/mm and a blaze angle of 44°, whereas three Cleartran prisms, used in double pass, act as cross-dispersers.

The Camera reimages the focal plane produced by the second pass through the collimator on the detector. It is a completely transmissive system, composed of three single lenses. The camera includes a filter, to cut transmissivity of the system above 2.0 µm. The detector is slightly tilted to correct for residual chromatic error.

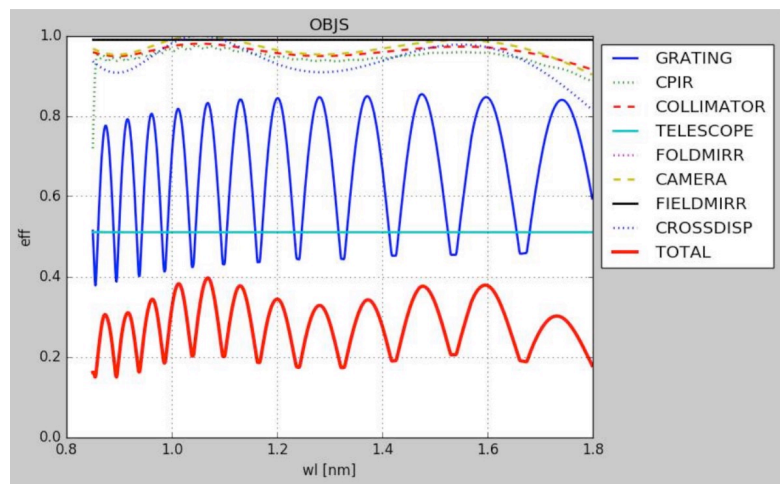

Figure 6. The total throughput of the NIR spectrograph, including the telescope.

The echelle spectral format is reported in Figure 5, showing that the complete spectrum fits well inside the 2k x2k detector. The inter-order gap is always greater than 10 pixels.

Throughput has been estimated using elements thickness-bulk-material-absorption (Zemax data) and estimates of the transmissivity and reflectivity of elements of telescope, CP and NIR spectrograph (Figure 6). The average throughput will be about 28%.

## 5. THE MECHANICS

The SOXS NIR spectrograph is composed by a D-shaped Vacuum Vessel (VV), a thermal shield and an optical bench, all in Aluminium [5]. The VV is interfaced with the derotator flange, via a set of large KM, that should support the 200 kg mass of the entire spectrograph. The optical bench is supported by 6 flexures made of high strength steel. An overall view of the spectrograph can be seen in Figure 7, together with a detail of the disperser unit.

On the wall of the VV a total of 7 entrances are welded: *i)* 3 DN 160 CF flange to ease the access to the detector, to mount the TMP, and to install a window behind the grating for alignment purpose, *ii)* 2 DN 50 ISO KF flanges to mount the feedthrough, *iii)* 2 DN 25 ISO KF flanges to mount the pressure gauges, *iv)* 1 DN 160 ISO K flange to mount the CCC system.

Removing the cover and the upper part of the thermal shield all the optical elements are shown together with the cryo-pump.

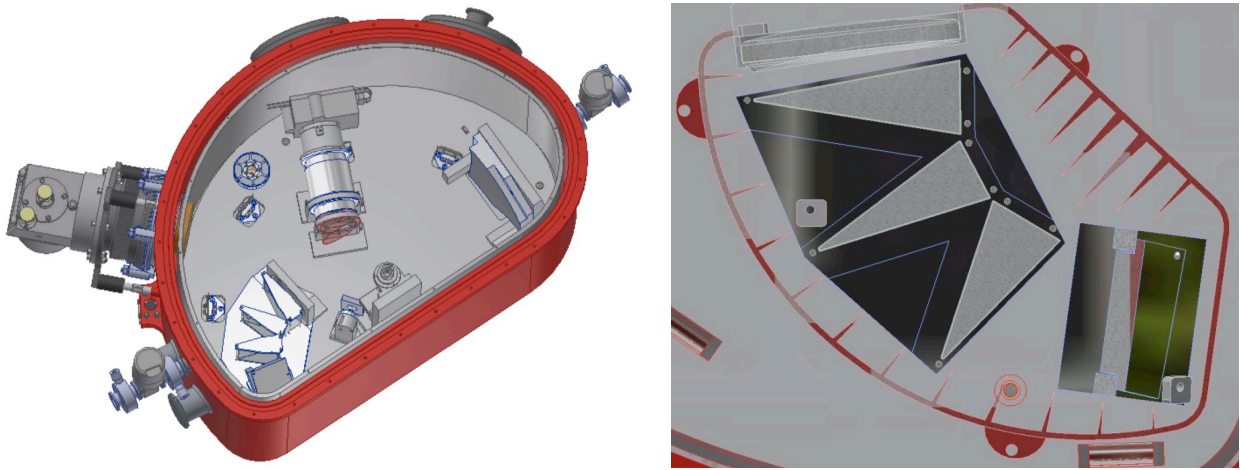

Figure 7. Left: an open view of the NIR cryostat vessel. The disperser section, the camera, the collimator and the (external) Cryo Compressor are clearly visible. Right: a detail of the disperser unit. The cross-disperser prisms and the grating are all encapsulated in a closed box, with a set of baffles to reduce the scattered light.

## 6. THE DETECTOR SYSTEM

The SOXS NIR detector system is composed by three main sub-systems: the detector, the electronic controller and the cryogenic assembly.

The detector is the HAWAII 2RG from Teledyne (Figure 8), with cut-off at 2.5 μm. The substrate removal technique extend the Q.E. respond down to 0.8 μm and beyond, allowing for a small and yet useful overlap with the UV-VIS wavelength range (Figure 9, left).

We decided to operate the array at 40 K, after a warning from ESO about the behavior of these arrays at temperatures below 80 K, where some of them shows a bump of persistence (as shown in Figure 9, right), that we want to avoid.

In Table 1 the main constructive parameters of the H2RG are shown, as reported by Teledyne.

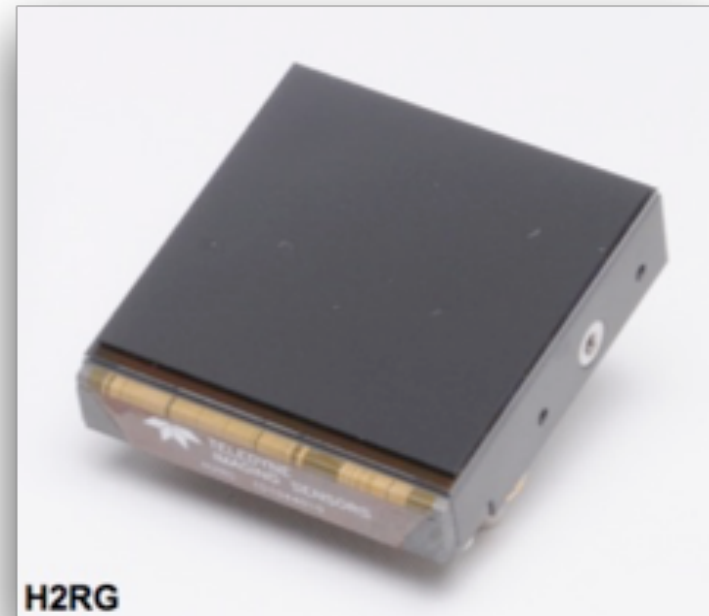

Figure 8. The Hawaii H2RG from Teledyne.

The detector will be controller via the new ESO NGC (New Generation Controller), result of ESO experience reached in many years, able to control both CCDs and NIR arrays.

The main NGC features are:
- High speed link
- Xilinx FPGA core elements on each board
- Digital parts and the sequencer are implemented in he FPGA
- Compact system
- Full compatibility with ESO software standard

The same controller has been adopted for the CCD of the UV-VIS arm of SOXS, allowing for a better spare and maintenance policy. In Figure 10, the general working scheme of the controller is shown.

The controller for the NIR array is composed by: a Transition module; a Backplane; two modules, one Basic Board that contains clock/bias generation and four video channels, and one additional Board

(AQ-32ch module) for video channels useful for NIR array which is equipped up to 32 reading channels (**Figure 11**).

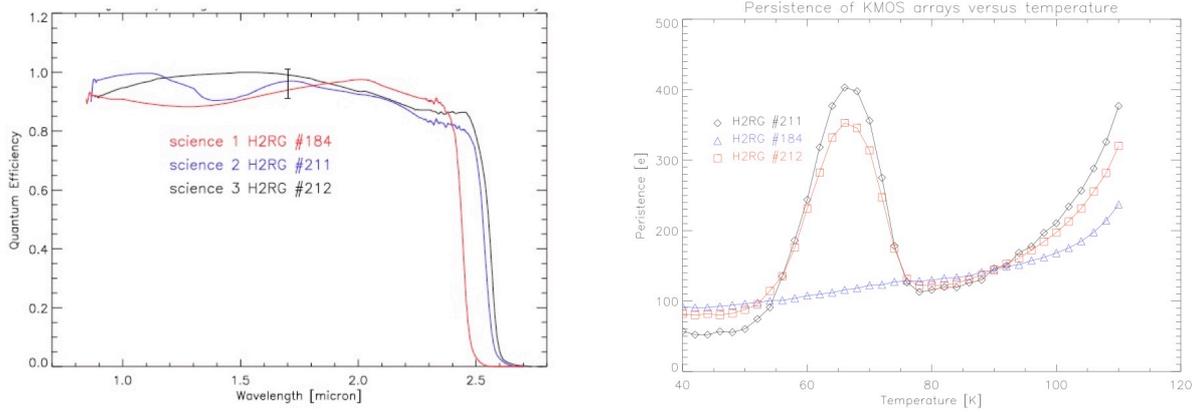

Figure 9. Left: the Q.E. of the H2RG arrays, treated with the substrate removal procedure. Right: persistence *vs* temperature in different HAWAII 2RG array.

Table 1. The main parameters for the H2RG array.

| Parameter | Value |
|---|---|
| Detector Material | MBE HgCdTe double layer planar heterostructures (DLPH) |
| Substrate | CdZnTe substrate, removed after hybridization to minimize fringing and optimize QE |
| Format | 2048x2048 |
| Cut-off wavelength, | $\lambda_c$ =2.5 µm |
| Pixel size | 18.0 µm |
| Number of outputs | 32 outputs for science frame access of internal bus outputs |
| Frame rate | 0.76 Hz using 32 parallel outputs |
| Reset by row | Non-destructive readout possible |
| Readout noise | Double correlated: < 20 $e^-$ rms<br>16 Fowler pairs < 7 $e^-$ rms |
| Storage capacity @ 1 V | 80 K$e^-$ |
| Dark current @ 40 K | < 0.005 $e^-$/s /pixel |

In order to prevent damage due to overheating, all housings shall be equipped with a thermal sensor, that can shut off the power to the NGC box.

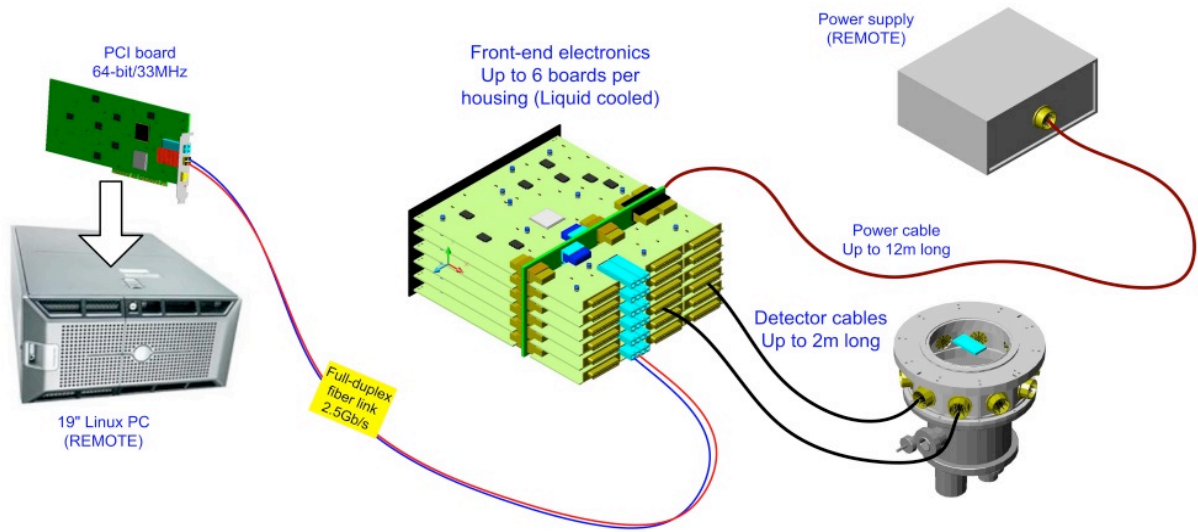

Figure 10. The general scheme of the ESO NGC controller.

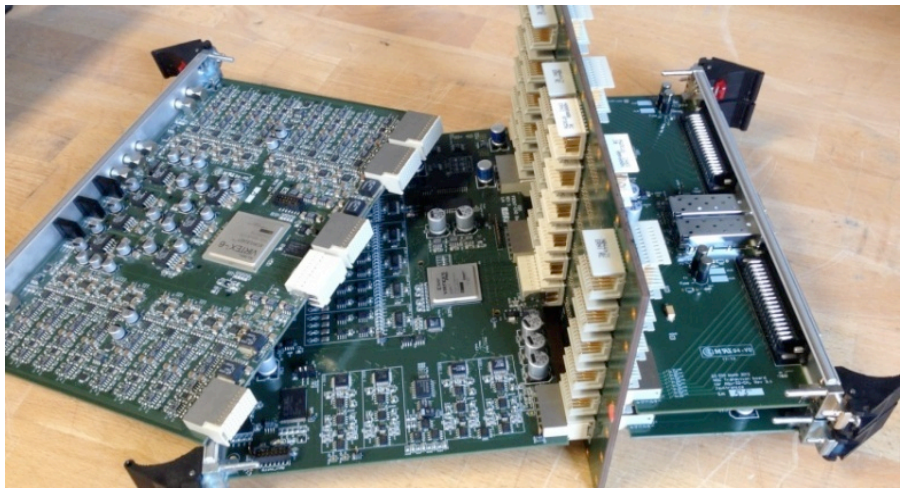

Figure 11. The NGC Controller: the lower module is the Basic Board, the second board is the AQ-32ch Module.

The SOXS-NGC controller is located very close to the instrument, because the cable between dewar and device is no longer than 2 m. Therefore we adopted a water-cooled housing to prevent any thermal dissipation into the telescope Nasmyth room.

The power supply for the NGC device is a 19-inch 3HE rack-mountable unit, which may be at a distance from the DFE of up to 12 meters.

The pre-amplifier (Figure 12) is located between the controller and the detector, close to the detector. For these reasons, this device is a cryogenic electronics. It is very compact and appear as an extension of the cryogenic detector cable.

An ESO-made PCI64 board (see Figure 13) with a fiber-optic connection to the controller is installed in the Data Acquisition Computer. The maximum theoretical bandwidth per interface is 256MB/s, which matches the 2.5 GBit/s fiber transmission rate. The bandwidth for actual data transmission is about 20% lower. With 1MHz ADC's, one 32-channel Acquisition Board will generate a maximum data rate of 64 MB/s.

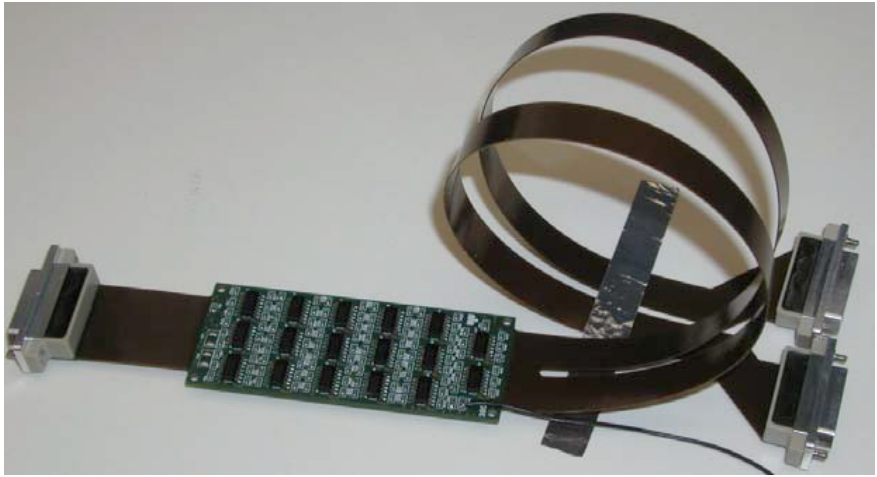

Figure 12. The cryogenic pre-amplifier, with cables.

For the detector housing we adopt the ESO standard for HAWAII 2RG 2Kx2K. The detector housing is a structure with three-legged system, as shown in **Figure 14**.

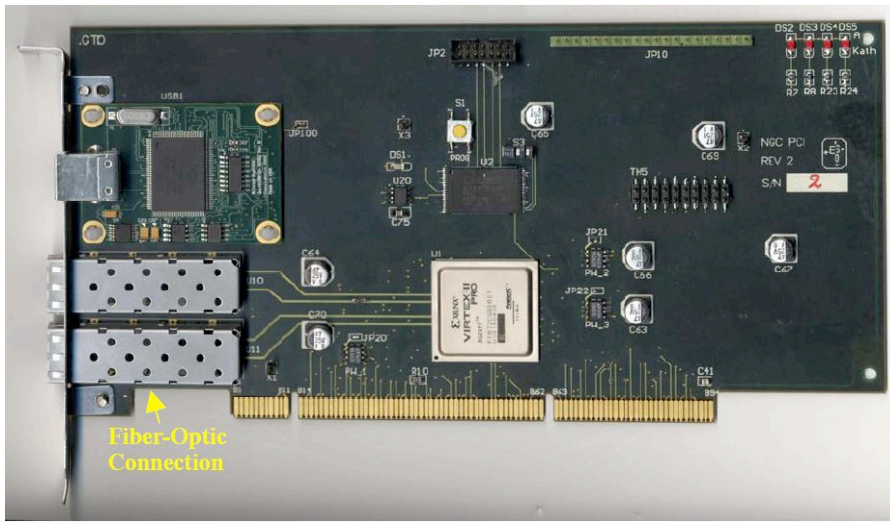

Figure 13. The PCI64 Board.

Electrical connection to the PGA is made using a flex circuit terminated in a 92 pin Hirose connector. To allow for a more robust interface, an additional flex cable connects to the HIROSE connector. On the other end a 72 pin Micro D connector is installed to the detector case. This setup allow us to disassembly independently the detector case from the pre-amplifier cable prevent any stress to the detector connector.

The mechanical detector mount (**Figure 14**) is mounted via three screws with spherical interface in V-shaped support parts, which allow for adjustment of the tilt and focus of the detector surface. The mount hosts the detector, the interface cables and the temperature sensor and heating resister. For the electrical connection to the preamp, a 72 pin Micro D connector is used. The electrical connection for the temperature control is done via a 9 pin Mirco D connector. The thermal connection is made with a copper braid to the backside of the detector mount.

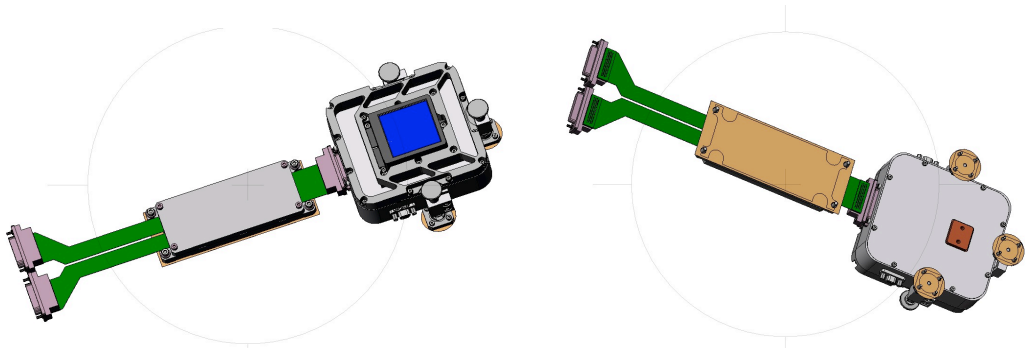

Figure 14. The detector unit (detector, holder, pre-amplifier and cables).

For the SOXS Control Electronics Design, see [6].

## 7. VACUUM & CRYO

The conceptual design of the Vacuum system for the NIR spectrograph is derived from the ESO standards (Vacuum and Cryogenics standard components, ESO–046147) and shown in Figure 15.

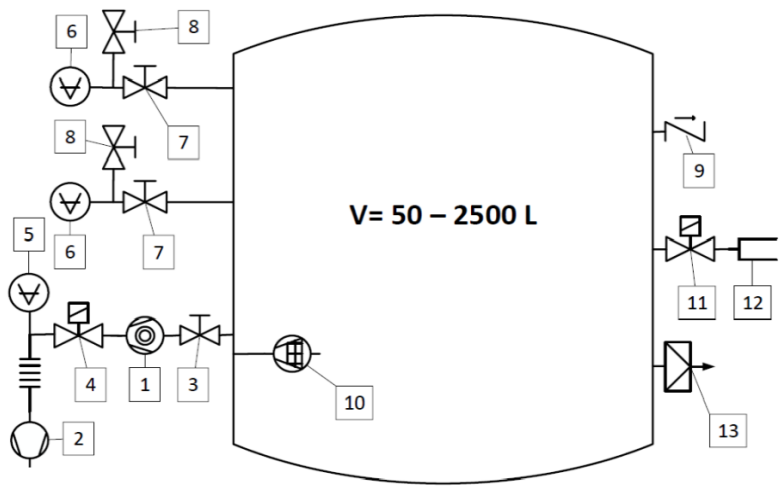

Figure 15. NIR Spectrograph Vacuum system.

The SOXS evacuation system is unique, in common between the two UV-VIS and NIR arms, but can operate independently for the two instruments. It consists of a turbo molecular pump (Pfeiffer Vacuum ATH 500M) and a separated pre-vacuum pump (Pfeiffer Vacuum ACP 28), both hosted on the platform at the Nasmyth focus of the NTT.

The two pumps are connected by a flexible bellow of the size DN 25, which shall be made as short as possible. Between the turbo pump and the pre-vacuum pump, an electromagnetic valve guarantees the insulation of the NIR cryostat. The NIR cryostat is equipped with two wide range vacuum gauges, each mounted at a separate port, in order to monitor permanently the residual vacuum pressure. One of the gauge is a full redundant device.

For controlled re-pressurization, the NIR cryostat is equipped with a venting valve and a gas connector as interface to the nitrogen gas supply. Finally, the NIR cryostat will be equipped with a sorption pump to guarantee the level of vacuum necessary for the operation of the NIR spectrograph.

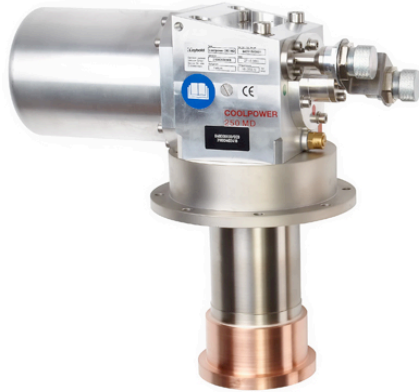

Figure 16. The Leybold CP 250 Md single Cold Head.

In the NIR spectrograph, detector, optical system (optics, mounts, bench…) thermal shield and mechanisms will be cooled down via a closed cycle cooler (CCC). We select a single-stage COOLPOWER 250 MD from Leybold (see Figure 16), which can be operated in any orientation. This cryo-cooler has the lowest vibration levels of its class and has the possibility of varying the motor frequency, that can be useful to avoid excitation of known eigenfrequencies of the telescope or other structures.

The 250 MD system comprises the following components: *i)* Cold head: COOLPOWER 250 MD with rotating flange DN 160 CF-F; *ii)* Compressor: COOLPAK 6000 HMD; *iii)* Flexible Helium lines.

The cold head will cool down the spectrograph optical bench, optics, thermal shield and mechanism down to T=150K, and the detector, the last lens and the filter at T=40K.

The cold head will be equipped with a custom made anti-vibration, to damp the vibrations introduced by a mechanical cooler to an acceptance level vibration specifications.

The temperature of the different parts of the spectrograph will be monitored through several temperature sensors, i.e. DT670 for the detector system and PT-103-AM for the spectrograph.

## 8. CONCLUSIONS

SOXS will pick up the heritage of X-Shooter and be the next tool for the study of transient events. It is envisaged to start observations in 2021. In this work we have presented the NIR arm of the SOXS spectrograph. For more detailed information about the several aspects of the NIR spectrograph, please refer to references [1] to [6].

## REFERENCES

[1] P. Schipani, et al., "The new SOXS instrument for the ESO NTT", Proc. SPIE 9908, 990841 (2016).
[2] P. Schipani et al., "SOXS: a wide band spectrograph to follow up the transients", Proc. SPIE 10702, 107020F (2018).
[3] R. Claudi, et al., "The Common Path of SOXS (Son of X-Shooter)", Proc. SPIE 10702, 107023T (2018).

[4] R. Zanmar Sanchez, et al., "Optical design of the SOXS spectrograph for ESO NTT", Proc. SPIE 10702, 1070227 (2018).
[5] M. Aliverti, et al., "The mechanical design of SOXS for the NTT", Proc. SPIE 10702, 1070231 (2018).
[6] G. Capasso, et al., "SOXS control electronics design", Proc. SPIE 10707, (2018).